\documentclass[aps,prl,twocolumn,tightenlines,superscriptaddress]{revtex4-1}
\usepackage{epsfig,rotating}
\usepackage{multirow}
\usepackage{amsbsy}
\usepackage{amsfonts}
\usepackage{stmaryrd} 
\usepackage{xfrac}   

\usepackage{lineno}

\begin{document}

\title{Water as a L\'{e}vy rotor}

\author{David\ A.\ Faux}
\affiliation{Department of Physics, University of Surrey,
	Guildford, Surrey GU2 7XH, United Kingdom}
\author{Arifah\ A.\ Rahaman}
\affiliation{Department of Physics, University of Surrey,
	Guildford, Surrey GU2 7XH, United Kingdom}
\author{Peter\ J.\ McDonald}
\affiliation{Department of Physics, University of Surrey,
	Guildford, Surrey GU2 7XH, United Kingdom}


   \date{\today}

\begin{abstract}
A probability density function describing the angular evolution of a fixed-length atom-atom vector as a L\'{e}vy rotor is derived containing just two dynamical parameters: the L\'{e}vy parameter $\alpha$ and a rotational time constant $\tau$.  A L\'{e}vy parameter $\alpha\!<\!2$ signals anomalous (non-Brownian) motion.  A molecular dynamics simulation of water at 298\,K validates the probability density function for the intra-molecular $^1$H--$^1$H dynamics of water.  The rotational dynamics of water is found to be approximately Brownian at sub-picosecond time intervals but  becomes increasingly anomalous at longer times due to hydrogen-bond breaking and reforming.  The rotational time constant lies in the range $8 \! < \! \tau \! < \! 11$\,ps.
The L\'{e}vy rotor model is used to estimate the intra-molecular contribution to the longitudinal nuclear-magnetic-resonance relaxation rate $R_{1,{\rm intra}}$ due to dipolar $^1$H--$^1$H interactions.  It is found that  $R_{1,{\rm intra}}$ contributes $65\,\pm 7$\% to the overall relaxation rate of water at room temperature.

\end{abstract}

\pacs{}
\maketitle

Water is a complex liquid and the most important substance in nature.  The water dynamics is critical to the understanding of protein folding,  bio-molecular water interface dynamics, water dynamics at the surfaces of porous materials, and the behavior of solvated ions, to list but a few examples. Each of these processes is fundamental to life itself.  

The dynamics of water may be probed by $^1$H nuclear magnetic resonance (NMR) techniques.  For instance, the longitudinal relaxation rate $R_1$ is exquisitely sensitive to the relative motion of pairs of $^1$H spins due to the collective effects of all water dynamical processes including vibration, bond stretching, libration,  tumbling, collisions and the breaking and reforming of H-bonds \cite{Abragam}.   Time constants associated with specific motions may be measured or inferred using infrared spectroscopies or isotope-enhanced NMR methods (see \cite{ropp2001rotational,qvist2012rotational,bakker2010vibrational,flamig2020nmr} and sources therein). 

The complex combination of translational and rotational motions can be sub-categorised into inter- and intra-molecular contributions.  The inter-molecular dynamics describes the evolution in space and time of a vector connecting a $^1$H spin on one molecule to a second $^1$H spin on a different molecule. The intra-molecular contribution accounts for pairs of spins on the same molecule.  The time-evolution of both the angular rotation and the vector length contribute to the observed relaxation rate.

This Letter focuses on intra-molecular $^1$H--$^1$H spin dynamics.  A description of the rotational dynamics of an intra-molecular $^1$H--$^1$H spin vector is presented which encapsulates the complex dynamics with just two parameters. The dynamics is described as a L\'{e}vy random walk \cite{levy1937}.  L\'{e}vy statistics have been used to characterise an enormous range of physical phenomena from foraging  albatrosses to entropy, from stock price fluctuations to heart beats \cite{tsallis1997levy}.  L\'{e}vy statistics applies to phenomena in which the more extreme events occur more frequently than predicted by a Brownian process leading to tails of the  probability density that follow a power law rather than an exponential decay.  This is often referred to as ``anomalous" diffusion. 

The L\'{e}vy model may be used to describe anomalous rotational diffusion in liquids in the bulk or confined in a porous medium, in large-molecular systems, and even in aqueous solutions containing ionic complexes.  Here, the L\'{e}vy rotor model is applied to liquid water.  Water is a relatively simple system that can be used to validate the L\'{e}vy model through molecular dynamics (MD) simulations using well-characterised force fields. Hydrogen-bond breaking and reforming is the known source of sudden angular changes leading to anomalous rotational dynamics. The L\'{e}vy model is then used to estimate the intra-molecular contribution to the spin-lattice relaxation rate $R_{1,{\rm intra}}/R_1$.  In pure water, $R_1$ is independent of the Larmor frequency and equal to 0.280\,s$^{-1}$\cite{krynicki1966proton}.  In ionic complexes, chemically-bound or physi-sorbed environments, $R_1$ is a function of Larmor frequency.

The simplest description of the intra-molecular rotational dynamics  dates from 1929 and the work of Debye \cite{debye1929polar} in which an atom-to-atom vector is described as a Brownian rotational diffuser executing a random sequence of small rotational steps.  The simple model provides a prediction for the NMR relaxation rate $R_1$  \cite{ Bloembergen.1948, Abragam}.  Ivanov \cite{ivanov1964theory} provided a general mathematical treatment of rotational diffusion and identified that the rotational time constants should be experimentally measurable.  A recent  examination of water dynamics undertaken by Laage and co-workers in a series of articles \cite{laage2006molecular,laage2006more,laage2007reorientional, laage2008molecular,laage2008residence} provides insight into the  dynamics not only of water but also of hydrated ions and complexes.
The reorientation dynamics of water due to hydrogen bonding is discussed \cite{laage2007reorientional}.  After a  H-bond is broken, the H-bond may reform with the same or a different oxygen acceptor.  Each process is characterised by a different time constant resulting in a highly-complex dynamical picture only partially clarified by experiment and MD simulations. Popov and co-workers  \cite{popov2016mechanism} provide a detailed review of H-bond network dynamics and its influence on particle dynamics, describing the role of processes (macroscopic motion, vibrational modes, libration) on multiple timescales up to 10\,THz.

\begin{figure}[tbh!]
	\unitlength1cm
	\begin{center}
		\includegraphics[width=0.42\textwidth, trim={0cm 16.5cm 0cm 2cm}]{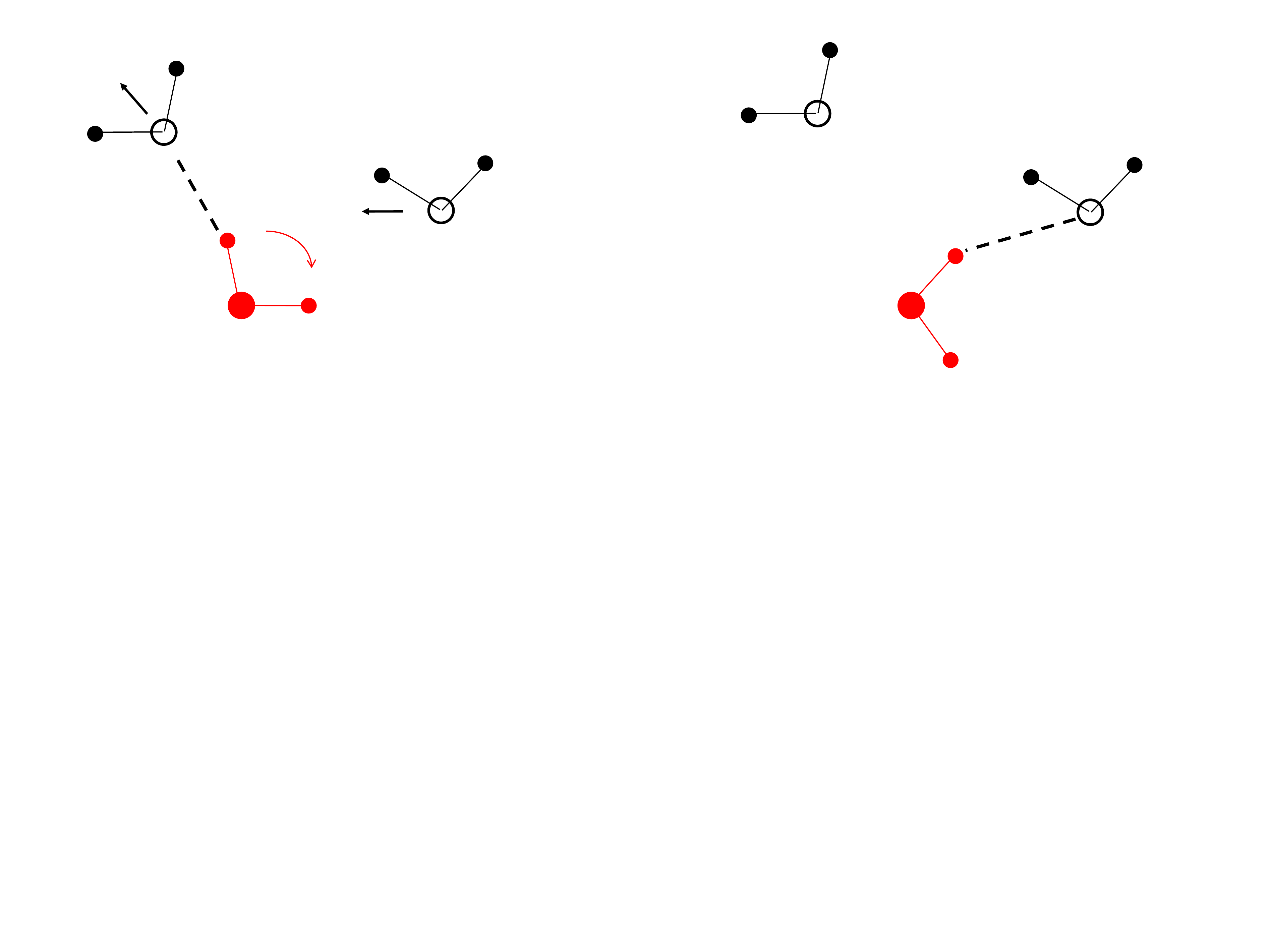}
		\includegraphics[width=0.42\textwidth, trim={2cm 3.5cm 2cm 1cm}]{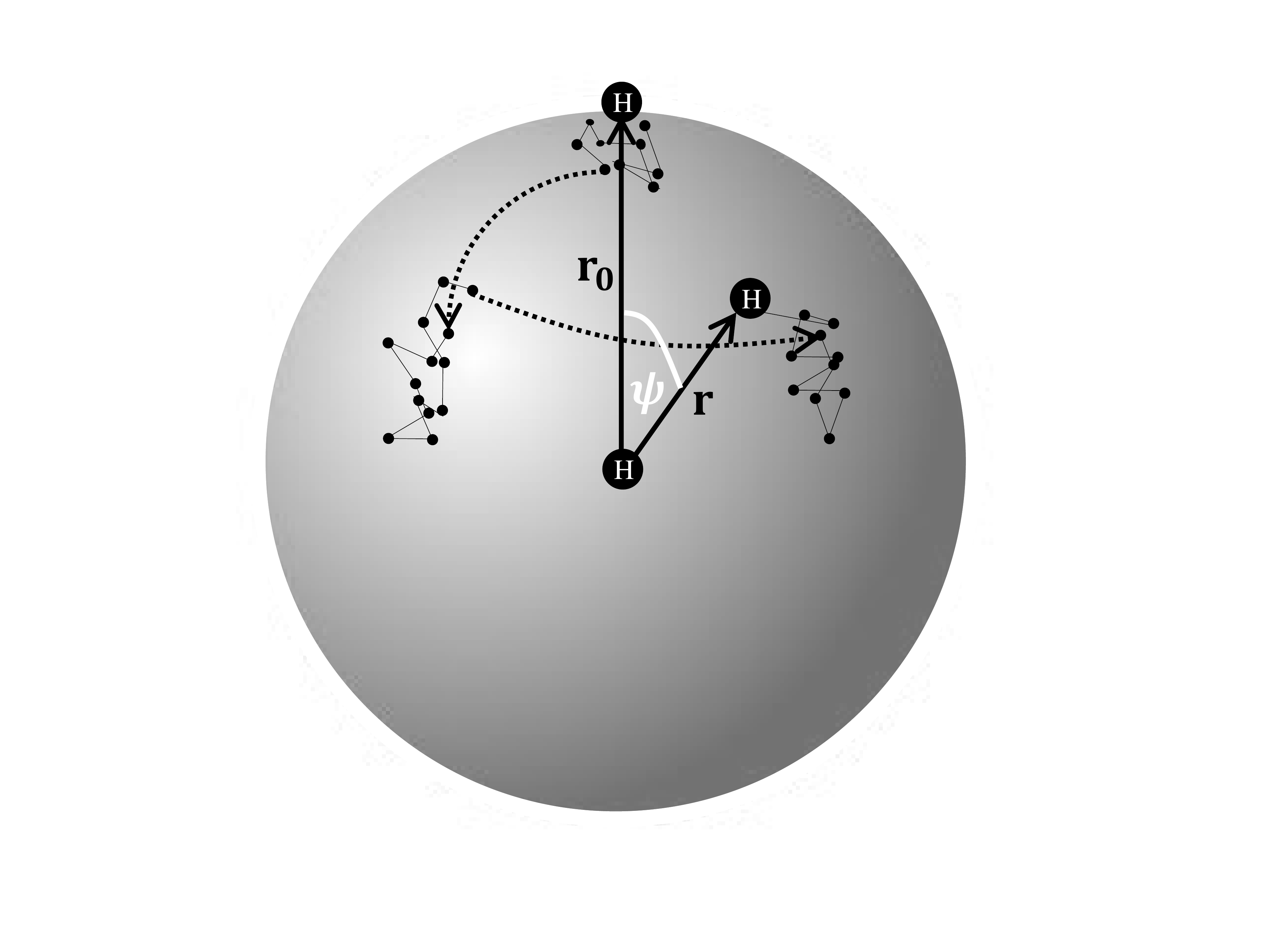}
		\caption{The L\'{e}vy model accommodates non-Brownian rotation dynamics such as H-bond breaking and reforming, illustrated top left by three water molecules with a H-bond (dashed line) which breaks and reforms with a different water molecule (top right).  Below, a coordinate system places one H-atom of a water molecule at the origin. The second H-atom is a fixed distance $a$ from the first and shown on the surface of a sphere of radius $a$.   The $^1$H--$^1$H vector at $t\!=\!0$ is labeled ${\bf r}_0$.  Rapid small angular movements of the $^1$H--$^1$H vector (solid lines) are punctuated by sudden large angular changes (dotted arrows). An example  $^1$H--$^1$H vector at time $t$ is labeled ${\bf r}$.  The smallest angle between ${\bf r}_0$ and ${\bf r}$ is $\psi$. \\[-0.9cm]} 
		\label{Fig1_Levy}
	\end{center}
\end{figure}

The L\'{e}vy model for a water molecule is illustrated in Fig.\,\ref{Fig1_Levy}.   The vector ${\bf r}_0$ connects the two H atoms in a single water molecule at time $t\!=\!0$ and is of fixed length $a$. The water molecule is in motion and the vector ${\bf r}$ changes direction (but not length) as a function of time. The angle $\psi$ is the smallest angle between the two vectors ${\bf r}_0$ and ${\bf r}$ at time $t$. Small fluctuations of angle $\psi$ may be described by Brownian dynamics. Occasionally the rotor undergoes a large change of angle which might be due to H-bond breaking and reforming with a different water molecule. The process then repeats with the angular evolution of ${\bf r}$ described as a L\'{e}vy rotor.  This picture is seen in MD simulations of supercooled liquids \cite{bohmer2001dynamics,qvist2012rotational} and in polymer systems\cite{tracht1999geometry}.  

The key result of this Letter is the derivation of a time-dependent angular probability density function $ P(\psi,t)$ of a fixed-length rotor that captures the anomalous diffusion as a L\'{e}vy random walk and which fully accounts for the angular boundary conditions; $\psi$ cannot take values less than zero or greater than $\pi$.  The derivation is presented as Supplementary Material.  The result is
\begin{eqnarray}
P(\psi,t) & = & N(t)  \left[ 1 +2 \sum_{p=1}^\infty e^{-p^\alpha t/\tau} \cos p\psi  \right] \label{eqn:PpsiL}
\end{eqnarray}
where $\tau$ is a characteristic rotational time constant and $N(t)$ is a normalisation constant (found in the Supplementary Material). The  L\'{e}vy parameter $\alpha$ is a measure of the departure from Brownian motion.  If $\alpha \!=\! 2$, Brownian rotational dynamics is recovered.  Anomalous diffusion is associated with $\alpha\! < \!2$ with $\alpha \!=\! 1$ a special case of Cauchy-Lorentz dynamics.  The rotational time constant $\tau$ in Eq.\,(\ref{eqn:PpsiL}) is defined differently to time constants bearing the same name in previous work.  Here, $\tau/2$ is the average time taken for a rotor to move through an angle of one radian.

The time-dependent probability density function of Eq.\,(\ref{eqn:PpsiL}) fully accounts for angular boundary conditions and has broad applicability. For example, $P(\psi,t)$ may be used to describe both Brownian or anomalous rotational motion of any vector connecting atom pairs in single molecules such as those containing paramagnetic markers for image enhancement in medical and biological research, or for rotors connecting aqueous ions and atoms in its first hydration shell.

The validity of the expression for $ P(\psi,t)$ was checked using MD simulation for water. All MD simulations rely on force fields that describe the intra-molecular and inter-molecular interactions between atom types. The flexible extended single-point charge (SPC/E) force field \cite{mark2001structure} was chosen because it has been widely used in previous MD studies supporting  NMR experimentation  \cite{ropp2001rotational,qvist2012rotational,calero20151h}. The MD model using the SPC/E force field incorporates fully-flexible bonds including angular vibration and bond stretching.
The LAMMPS package was used for all simulations \cite{LAMMPS}.  A system of 13500 water molecules in a cubic box was pre-equilibrated to a temperature of 298\,K and zero pressure under isothermal and isobaric conditions.  The simulation temperature was controlled using the Hoover thermostat \cite{Hoover.1985} and periodic boundary conditions were applied.   The data gathering used the NVT (constant number of particle, volume and temperature) ensemble for a time period of 100\,ps with atomic positions recorded at 1\,ps intervals.

\begin{figure}[tbh!]
	\unitlength1cm
	\begin{center}
		\includegraphics[width=0.5\textwidth, trim={1cm 9cm 0cm 10cm}]{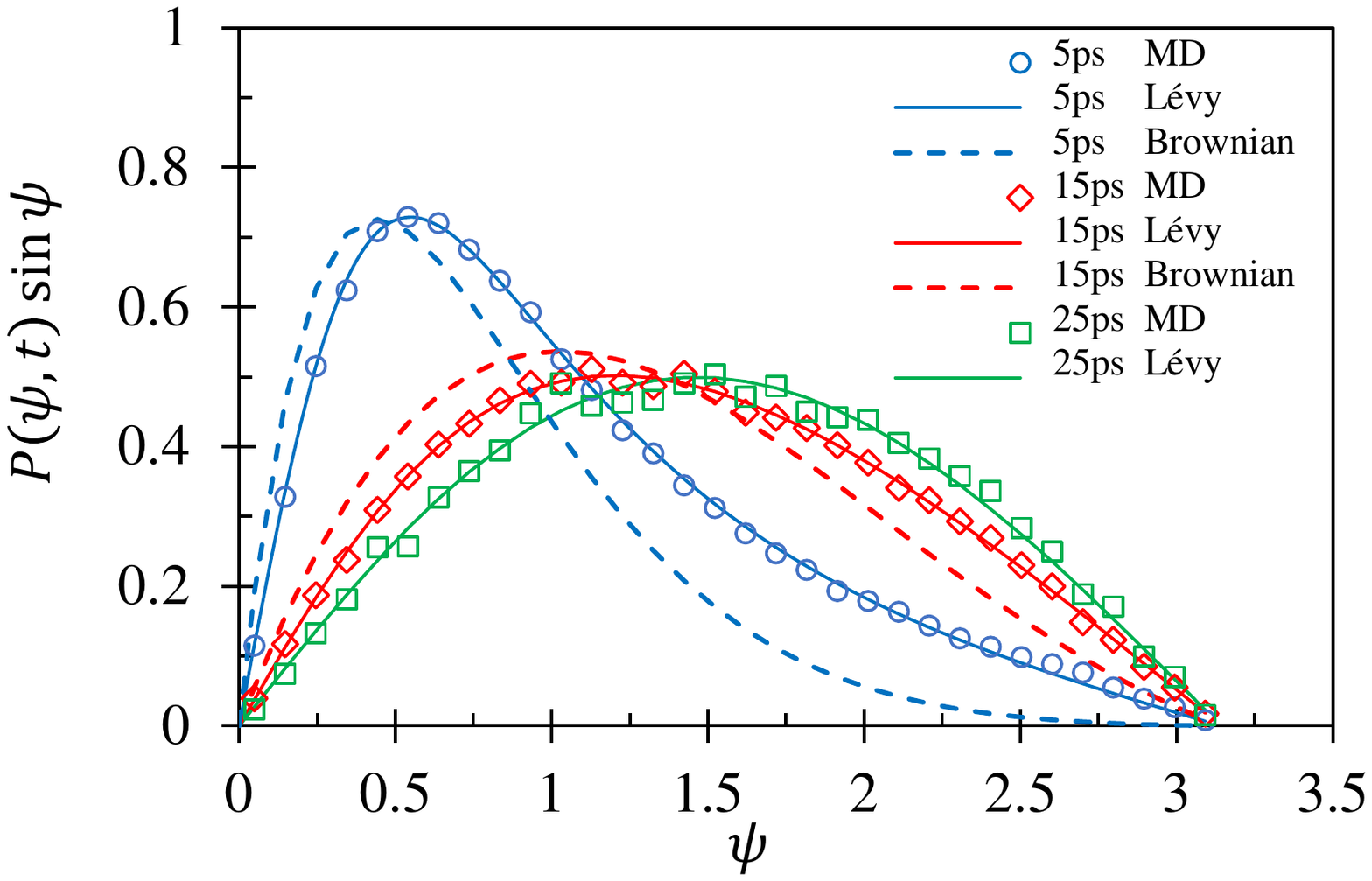}
		\caption{The figure presents the probability density function $P(\psi,t) \sin \psi$ obtained from MD simulations (symbols).   The L\'{e}vy model (solid lines) and Brownian models (dashed lines: 5\,ps and 15\,ps only for clarity) are obtained using Eq.\,(\ref{eqn:PpsiL}). \\[-1cm]}
		\label{Fig2_Ppsi}
	\end{center}
\end{figure}
The probability density function $P(\psi,t)\sin \psi$ is computed for times $t\!=\!0.01-25$\,ps using Eq.\,(\ref{eqn:PpsiL}) and fit to the distribution obtained from MD simulation.  The optimum fit as judged by minimising the least-squared deviation to provide $\alpha$ and $\tau$.  Figure\,\ref{Fig2_Ppsi} compares  the evolutions at $t\!=\!5, 15$ and $25$\,ps.  The distributions for a Brownian rotor provide a poor match to the MD results. By contrast, the L\'{e}vy model of Eq.\,(\ref{eqn:PpsiL}) provides excellent fits confirming that the rotational dynamics of water is anomalous.   

\begin{figure}[tbh!]
	\unitlength1cm
	\begin{center}
		\includegraphics[width=0.50\textwidth, trim={1cm 9cm 0cm 10cm}]{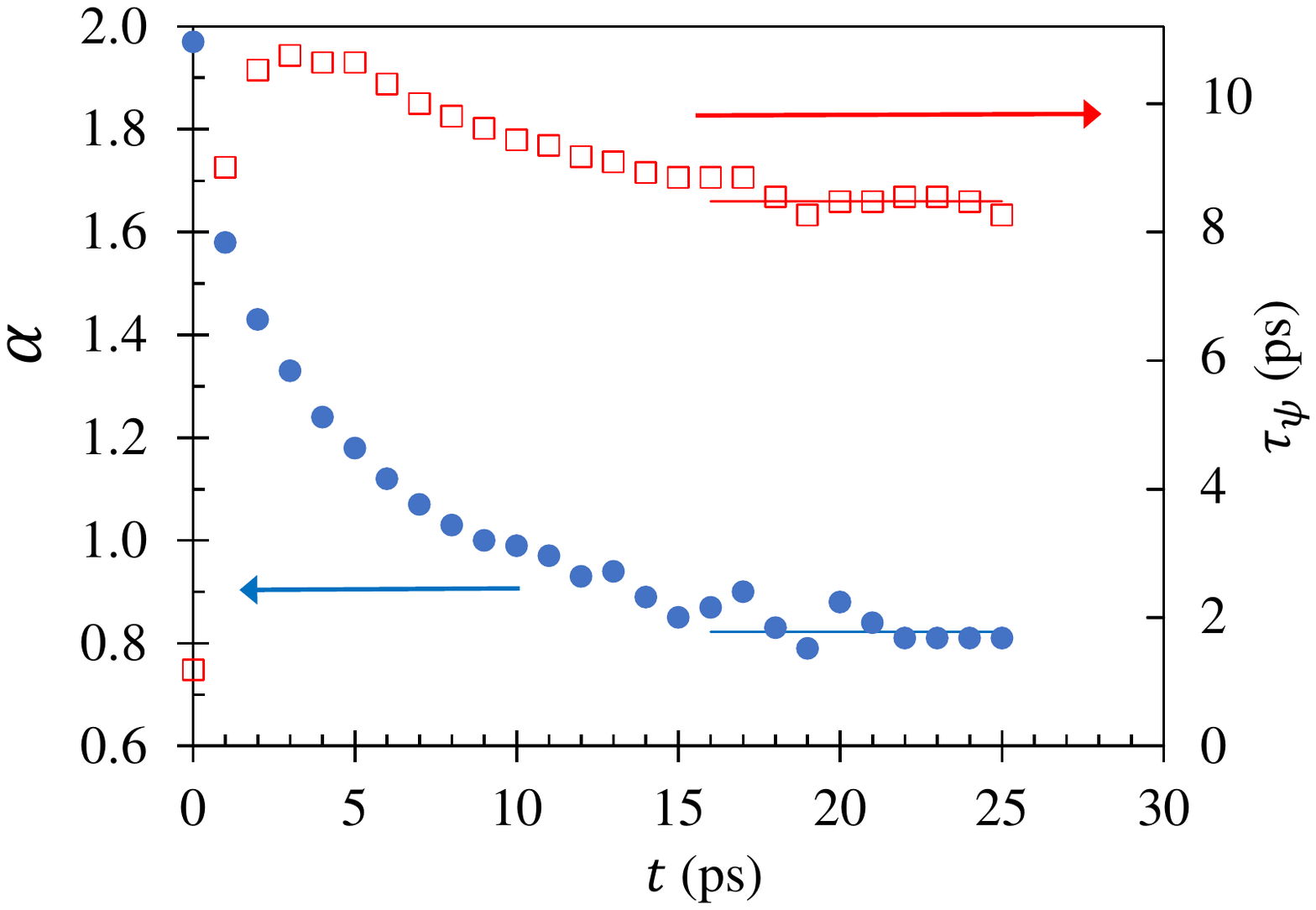}
		\caption{The figure presents the L\'{e}vy parameter $\alpha$ and rotational time constant $\tau$ obtained from fits to MD results using Eq.\,(\ref{eqn:PpsiL}) for time intervals $t\!=\!0.01$\,ps and 1--25\,ps.\\[-7mm]} 
		\label{Fig3_alpha_tau}
	\end{center}
\end{figure}

The results for $\alpha$ and $\tau$ are presented in Fig.\,\ref{Fig3_alpha_tau}. As anticipated,  $\alpha \! \rightarrow \! 2$ at very short time intervals indicating Brownian rotational dynamics with small angular changes characterised by a picosecond time constant.  The  rotational motion becomes increasingly anomalous for longer time intervals due to large changes in angle associated with H-bond breaking and reforming causing $\alpha$ to decrease. $\alpha$ attains a constant value of approximately 0.8 for time intervals in excess of 15\,ps.  The mean values of $\alpha$ and $\tau$ were found from time intervals 18\,ps to 25\,ps inclusive to be $\bar{\alpha}\!=\!0.82\pm 0.12$ and $\bar{\tau}\!=\!8.5\pm 0.4$\,ps with 95\% confidence. 

Infrared (IR) spectroscopy may be used to measure rotational time constants in water.  An O--H bond switching time of approximately 2.5\,ps is suggested from both experiment and MD  \cite{bakker2010vibrational}.  Ultrafast 2D IR vibrational-echo chemical-exchange spectroscopy has been used to measure an anion-oxygen H-bond switching time constant of 6.5\,ps  \cite{moilanen2009ion}.  The rotational time constant $\bar{\tau}/2\approx 4$\,ps found here corresponds to the characteristic time for an angular change of 1 radian, approximately the H-bond switching angle, and is consistent with the experimental values.

Experimental $^1$H NMR relaxometry of water measures $R_1$  but cannot resolve the individual dynamic components that contribute to the relaxation rate.  
In a model, however, the contributions of the intra-molecular and inter-molecular components may be separately estimated.  The intra-molecular contribution is associated with the rotation of the fixed-length $^1$H--$^1$H vector in the applied magnetic field and the inter-molecular contribution is due to the relative translational motion of pairs of $^1$H spins on different molecules and includes changes in both vector length and angle.
The rotational time constant $\tau$ presented in Fig.\,\ref{Fig3_alpha_tau} as a function of time interval characterises the average intra-molecular $^1$H--$^1$H bond rotation due to all dynamical processes.  

Equation (\ref{eqn:PpsiL}) supplies an excellent description of the rotational dynamics of any vector {\bf r} of fixed length $a$ connecting two hydrogen nuclei in water as a L\'{e}vy rotor in terms of just two parameters, $\alpha$ and $\tau$.  Equation (\ref{eqn:PpsiL}) is used to estimate the intra-molecular contribution to the longitudinal relaxation rate $R_1$ of pure water.  The room temperature relaxation rate for pure water was studied extensively by $^1$H NMR relaxometry from the 1950's to the 1970's.  Consistent values of $R_1$ were achieved once the contribution of dissolved oxygen had been identified and removed.
Krynicki \cite{krynicki1966proton} summarised the collective data for air-free pure water at room temperature. The  room-temperature relaxation rate interpolated from the figure 1 is  $R_1=0.280 \pm 0.010$\,s$^{-1}$.  

The L\'{e}vy model for intra-molecular rotational assumes that the distance between the two $^1$H spins on the same water molecule is fixed and that the rotational evolution is described by the probability density function  $P(\psi,t)$ given by Eq.\,(\ref{eqn:PpsiL}).  The time-dependent dipolar correlation function $G(t)$ describes how spin-pairs move relative to each other and is derived in the Supplementary Material as Eq.\,(S16) as \cite{Messiah.1965,Sholl1974nuclear}
\begin{eqnarray}
\!\!\!\!\!\!G(t) & \!=\! &  \frac{1}{a^6} \!\!\int_0^\pi \!\! P_2 ( \cos \! \psi )   P(\psi,t)  
\sin  \psi \, d \psi  \label{eqn:G3App}
\end{eqnarray}
where $P_2(x)$ is the second-rank Legendre polynomial, and $a$ is the intra-molecular $^1$H--$^1$H distance.  Critical to an accurate determination of $R_1$ is the choice of $a$.  Sawyer has proposed that, for NMR applications, $a\!=\! \langle r_{\rm HH}^{-3} \rangle^{-1/3}$ recognising the $r^{-3}$ distance dependence of the dipolar interaction\cite{Sawyer}.  A summary discussion in the  Supplementary Material justifies $a\!=\!0.1545\pm0.0007$\,nm.

The spectral density function $J(\omega)$ is the Fourier transform of the dipolar correlation function,
\begin{eqnarray}
J(\omega) & = & 2 \int^\infty_{0} \!\!  G(t) \: \cos \omega t \: dt, \label{Jeqn}
\end{eqnarray}
and the longitudinal relaxation rate  is expressed as \cite{Abragam}
\begin{eqnarray}
R_1 & = &  \frac{1}{5} \psi_{I\!I} \left[  J(\omega) + 4 J(2 \omega)\right] \label{T1f} 
\end{eqnarray}
where  $\omega$ is the Larmor frequency of a $^1$H spin in the applied static field, $ \psi_{I\!I}\!=\! \left( \mu_0 / 4\pi \right)^2 \gamma_{\rm I}^4 \:\hbar^2 I(I+1)$,  $\gamma_{\rm I}$ is the proton gyromagnetic ratio and $I\!=\!\frac{1}{2}$. For water $\psi_{I\!I}\!=\!4.275 \times 10^{11}$\,m$^6$\,s$^{-2}$.  

The spin-lattice relaxation rate $R_1$ is computed directly from the values of $\alpha$ and $\tau$.  It is sufficient to compute the correlation function $G(t)$ of Eq.\,(\ref{eqn:G3App}) for the range 1--25\,ps at 1\,ps intervals with $G(t)$ at intermediate times obtained by quadratic interpolation.  Contributions beyond 25\,ps are negligible indicating that only dynamics with time constants less than 25\,ps contribute to the relaxation rate in water (dynamic equilibrium has been attained for $t\! \geq \!25$\,ps and $P(\psi,t)$ is unchanging). Therefore, the numerical values of $\alpha$ and $\tau$ presented in Fig.\,\ref{Fig3_alpha_tau} for times 1--25\,ps may be used directly.  $R_{1,{\rm intra}}$ is computed using Eqs.\,(\ref{eqn:G3App})--(\ref{T1f}).    

We find $R_{1,{\rm intra}} = 0.183$\,s$^{-1}$ and is independent of frequency over the range 1\,kHz--40\,MHz. The experimental value of $R_1$ is 0.280\,s$^{-1}$ at 28\,MHz and at room temperature  \cite{krynicki1966proton} so that $R_{1,{\rm intra}}/R_1\!=\!0.65$.
The uncertainties in the dynamical time constants and their impact on $R_{1,{\rm intra}}$ are assessed in the Supplementary Material.  Time constants for specific motions may differ from experiment by up to 20\% but the single $\tau$ used here encompasses all dynamics that contribute over a time interval. It is argued that a reasonable estimate of the overall uncertainty of the flexible SPC/E force field to reproduce $\tau$ is $\pm 10$\%.  The value of $\tau$ in Fig.\,\ref{Fig3_alpha_tau} is changed by $\pm$10\% to obtain $R_{1,{\rm intra}}/R_1\!=\!0.65\pm0.07$. 

The first estimation of the intra-molecular contribution to $R_1$ in the 1960's treated a molecule as a rotating sphere in a viscous fluid concluding that the intra-molecular contribution was approximately 50\% of the total rate \cite{favro1960theory, Abragam}.  The most recent comparable work is by Calero and co-workers \cite{calero20151h} who determined $R_1$ from MD simulations of water using four different intra- and interatomic force fields.  Force field TIP4P/2005 yielded the best estimate of $R_1$ at room temperature for which $R_{1,{\rm intra}}/R_1\!=\!0.67$.  Calero {\it et al} determine both $R_{1,{\rm inter}}$ and the diffusion coefficient from MD simulation.  Their diffusion coefficient is 13\% lower than the experimental value\cite{krynicki1978pressure}. It is calculated in the Supplementary Material that, if their MD simulations had yielded the experimental diffusion coefficient, this would yield $R_{1,{\rm intra}}/R_1\!=\!0.65$.  Calero's overall value for $R_1$ is  6\% lower than the experimental value.  The relaxation rate scales as $a^{-6}$ and this remaining error is probably associated with the quality of the TIP4P/2005 force field to reproduce the $^1$H--$^1$H distance.

In summary,  the key result is Eq.\,(\ref{eqn:PpsiL}) which provides the first description of an anomalous (L\'{e}vy) rotor.  Equation\,(\ref{eqn:PpsiL}) may be used to model rotational dynamics of pairs of atoms in molecular liquids, molecules with restriction motion such as bio-molecules or proteins, and aqueous ionic complexes.
Equation\,(\ref{eqn:PpsiL}) is shown to provide an excellent description of the rotational dynamics of the intra-molecular $^1$H--$^1$H vector in water.  Water is found to be a highly-anomalous diffuser due to H-bond breaking and reforming characterised by  a long-time mean L\'{e}vy parameter $\bar{\alpha}\!=\!0.82\pm 0.12$.  The headline values for $R_{1,{\rm intra}}/R_1$ obtained here and from Calero {\it et al} (corrected) are identical, despite different methods of calculation, and found to be $65 \pm 7$\%.

This work has received funding from the European Union Horizon 2020 Research and Innovation Programme under the Marie Sk\l odowska-Curie Innovative Training Networks programme grant agreement No. 764691.

\bibliography{References_master_2021e}

\end{document}



\begin{center}
\bf\LARGE{Water as a L\'{e}vy rotor}

{\bf\Large Supplementary material} 
\end{center}

\renewcommand{\thefigure}{S\arabic{figure}}
\setcounter{figure}{0}

\section{The probability density function for a L\'{e}vy rotor}

The objective is to derive a general expression for the angular probability density function $P(\psi,t)$ which describes the anomalous diffusion of a fixed-length rotor as illustrated in Fig.\,S1.


\begin{figure}[tbh!] 
	\unitlength1cm
	\begin{center}
		\includegraphics[width=0.7\textwidth, trim={2cm 12cm 2cm 1cm}]{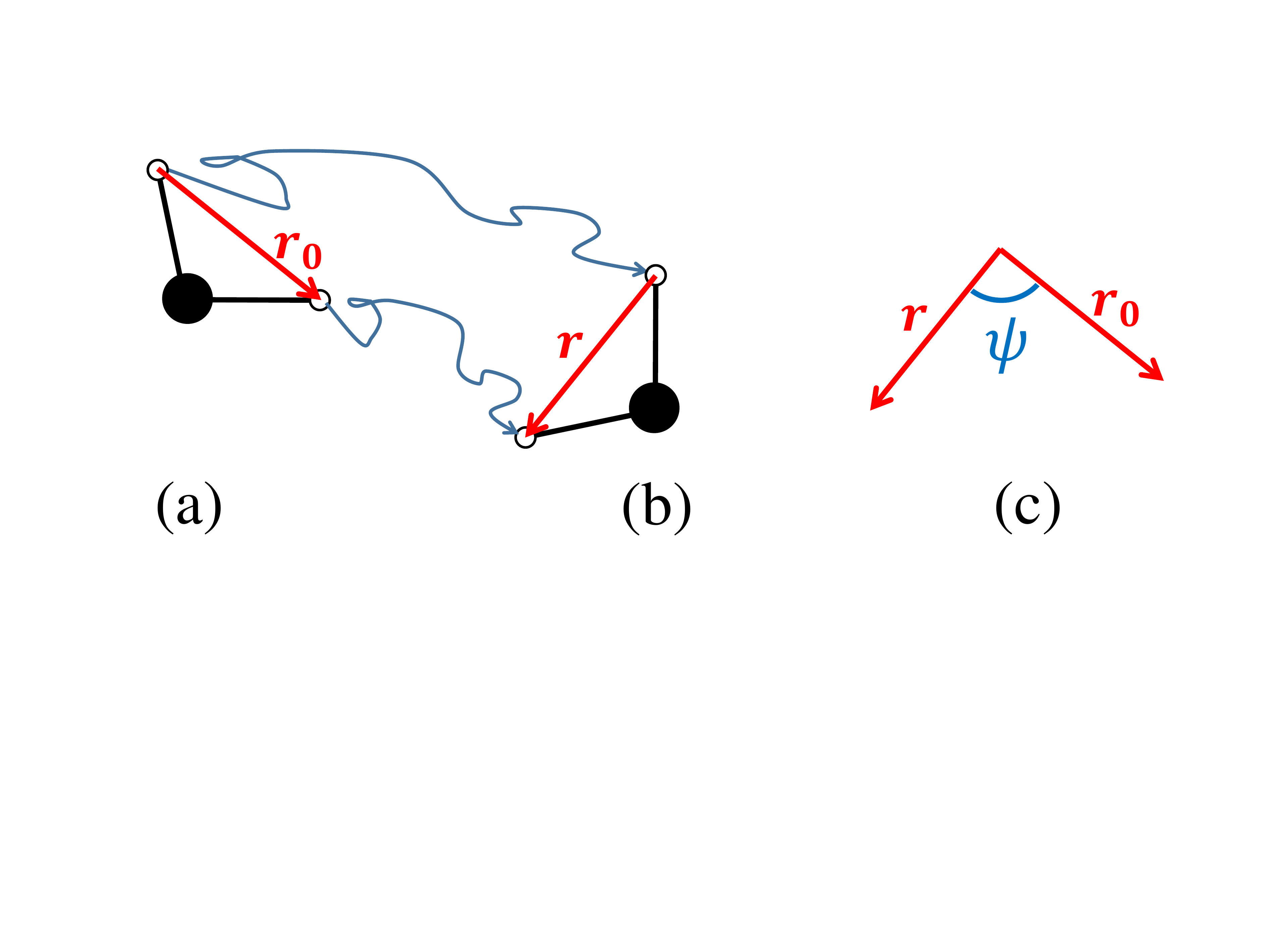}
		\caption{In (a), a water molecule is positioned at $t=0$ with  hydrogen spins linked by a vector $\bf{r}_0$. In (b) the molecule moves for a time $t$ and $\bf{r}$ links the same two hydrogen spins. $\psi$ is the smallest angle between the two vectors as shown in (c).}
		\label{FigS1intra_schematic}
	\end{center}
\end{figure}

 In Fig.\,S1(a), at $t=0$, a pair of water protons are connected  by the vector ${\bf r}_0$.  The molecule undergoes rotational and translational motion and, at time $t$, the same two protons are separated by a vector ${\bf r}$. Note that $|{\bf r}|\,=\,|{\bf r}_0|$ always.  The change in angle of the two vectors is $\psi$ as shown in Fig.\,S1(c).
 
 Initially, some standard results for one-dimensional (1D) diffusion of a single particle are reviewed. The probability density $p(x,t\,|x_0)$ describing the motion of a particle in 1D located at $x_0$ at $t\!=\!0$ is the solution to the Fokker–Planck diffusion equation $\partial p/\partial t \!=\! D \partial^2 p/\partial x^2$ as
 \[
 p(x,t\,|x_0) = \frac{1}{(8 \pi D t)^{1/2}}  e^{-(x-x_0)^2/8Dt} \tag{S1} \label{eqn:1Dxx0}
 \]
 where $D$ is the diffusion coefficient.
 Shortly, $x$ will be replaced by angle $\psi$.  It is evident from Fig.\,S1 that $\psi_0\!=\!0$ always, and so henceforth $x_0\!=\!0$. At time $t$,
  \[
 p(x,t) = \frac{1}{(8 \pi D t)^{1/2}}  e^{-x^2/8Dt}. \tag{S2} \label{eqn:1Dx}
 \]
 The distribution in space is now examined at time $t$ with $a\!=\!2Dt$ so that
   \[
 p(x) \:=\: \frac{1}{2(a \pi)^{1/2}}  e^{-x^2/4a} \:=\: \frac{1}{2 \pi}\int^\infty_\infty e^{-ak^2} \, \cos kx \, dk \tag{S3} \label{eqn:1DxFT}
 \]
where the second expression presents the Fourier transform of the Gaussian function $p(x)$.   It is noted that a stable symmetric L\'{e}vy distribution centered at $x=0$ may be written
    \[
 L_\alpha(x) \:=\: \frac{1}{2 \pi}\int^\infty_\infty e^{-a |k|^\alpha} \, \cos kx \, dk\tag{S4} \label{eqn:Levy}
 \]
 where $\alpha$ is the L\'{e}vy parameter used in main article \cite{levy1937}.  Clearly $p(x)\!=\! L_2(x)$, an equation that is exploited later.
 
The particle is now confined to a half-space $0 \! \leq x \leq h$. The Fokker–Planck diffusion equation must be solved using the boundary conditions that $P(x,t)\!=\!0$ and $\partial P(x,t)/ \partial x \!=\!0$ at $x\!=\!0,h$.  This calculation is presented by Carslaw \cite{carslaw1959conduction} and reproduced by Bickel \cite{Bickel.2007} to obtain
\[
p_c(x,t) =  \frac{1}{h} \left[ 1 +2 \sum_{p=1}^\infty e^{-D p^2 \pi^2 t / h^2} \cos \frac{p \pi x}{h} \right]. \tag{S5} \label{eqn:PxConfined}
\]
The boundary conditions unsurprisingly lead to a series solution expression as the sum of Fourier terms.  Equation (\ref{eqn:PxConfined}) is normalised provided the {\it a priori} probability density of a particle being found at position $x$ within $0 \! \leq x \leq h$ is uniform.
 
Attention is now turned to the evolution of the angle $\psi$. Simple models describing the angular time evolution treat $\psi$ as executing a Brownian random walk of small angular steps.  Such models are limited to short times because $\psi$ can become arbitrarily large whereas $\psi$ is bounded; $\psi$ cannot take values less than zero or greater than $\pi$.  An expression for  the time-dependent probability density function $P(\psi,t)$ that describes the evolution of the angle $\psi$ subject to the  conditions $\psi\!=\!0$ at $t\!=\!0$ and $0 \leq \psi \leq \pi$ may be found by adapting Eq.\,(\ref{eqn:PxConfined}) to the bounded rotor by changing the coordinate $x$ to angle $\psi$ and $h$ to $\pi$ to yield
\[
P_B(\psi,t) = N(t)  \left[ 1 +2 \sum_{p=1}^\infty e^{-p^2t/\tau} \cos p\psi  \right]. \tag{S6} \label{eqn:PpsiBrownian}
\]
The 1D diffusion constant $D$ becomes $\tau^{-1}$ where $\tau$ is a rotational time constant.  $\tau/2$ represents the mean time for a rotor to move through an angle of one radian.  The subscript B on the angular probability density function affirms Eq.\,(\ref{eqn:PpsiBrownian}) as applicable to Brownian rotors.

In contrast to the linear system, the {\it a priori} probability density of a rotor moving through an angle $\psi$ is not uniform.  If two vectors are placed at random from the center of a sphere to its surface, the angular separation is distributed as $\sin \psi$. The transformation from variable $x$ to angle $\psi$ therefore necessitates a time-dependent normalisation constant $N(t)$ which ensures that
\[
\int_0^\pi P(\psi,t) \sin \psi \, d\psi = 1 \tag{S7} \label{eqn:psiNorm}
\]
for all $t$.  The normalisation term $N(t)$ is 
\[
N(t)  = \frac{1}{2} \left[ 1 - 2 \!\! \sum_{p=2,4...}^\infty   \!\! \frac{e^{-p^2 t/\tau}}{(p^2-1)} \right]^{-1}. \tag{S8}  \label{eqn:NtB}
\]

Equation (\ref{eqn:PpsiBrownian}) provides the rotational probability density function for a Brownian rotor including angular boundary conditions in terms of a rotational time constant $\tau$.  The summation in Eq.\,(\ref{eqn:PpsiBrownian}) is a Fourier series that mirrors the Fourier integral of Eq.\,(\ref{eqn:1DxFT}) with the summation index $p$ taking the role of the Fourier variable $k$.  The L\'{e}vy distribution Eq.\,(\ref{eqn:Levy}) differs from the Fourier integral representation of the Gaussian distribution of Eq.\,(\ref{eqn:1DxFT}) only by the L\'{e}vy parameter $\alpha$ which replaces the ``2" on the Fourier variable.  It follows therefore that 
anomalous (L\'{e}vy) diffusion is described by Eq.\,(\ref{eqn:PpsiBrownian}) by replacing the ``2" on the series index $p$ by the L\'{e}vy coefficient $\alpha$, thus
\[
P_L(\psi,t) = N(t)  \left[ 1 +2 \sum_{p=1}^\infty e^{-p^\alpha t/\tau} \cos p\psi  \right] . \tag{S9} \label{eqn:PpsiBrownianLevy}
\]
Finally, the normalisation term $N(t)$ becomes 
\[
N(t)  = \frac{1}{2} \left[ 1 - 2 \!\! \sum_{p=2,4...}^\infty   \!\! \frac{e^{-p^\alpha t/\tau}}{(p^2-1)} \right]^{-1}. \tag{S10}  \label{eqn:Nt}
\]
Equation (\ref{eqn:PpsiBrownianLevy}) is the key result.   This expression can be used to describe the anomalous rotational diffusion of any molecular bond or, indeed, of any vector connecting one atom to another in a liquid.

\section{The dipolar correlation function for a L\'{e}vy rotor}

The nuclear magnetic resonance (NMR) relaxation rate measured from water includes a contribution due to the rotational motion of pairs of intra-molecular proton spins.  
The key to developing a model describing the frequency dependence of the NMR relaxation rate is the determination of the time-dependent dipolar correlation function $G(t)$.   $G(t)$ captures all the relevant dynamical information describing how pairs of spins move relative to each other.  In any sample with randomly-oriented fluid spaces, $G(t)$ is determined from  \cite{Messiah.1965,Sholl1974nuclear}
\[
G(t) =       \left< \frac{ P_2( \cos \psi )  }{r_0^3 r^3} \right> \tag{S11} \label{eqn:Gt_ave}
\]
where $P_2(x)=\tfrac{1}{2}(3x^2-1)$ is the second-rank Legendre polynomial. Introducing a space- and time-dependent probability density function yields an equivalent expression 
\[
G(t) =  \int_{\mathbb{R}^3} \!\!  \int_{\mathbb{R}^3_0} \frac{ P_2( \cos \psi )}{r_0^3 \: r^3}  P({\bf r},t  \cap {\bf r}_0)  \: d^3 {\bf r}_0 \, d^3 {\bf r}, \tag{S12} \label{eqn:Gt_integral}
\]
where $P({\bf r},t \cap \:{\bf r}_0)$ is the probability density function describing the probability distribution of pairs of spins separated by  ${\bf r}_0$ at $t=0$ {\em and} by ${\bf r}$ at time $t$. The subscript 0 on all quantities indicates the value at $t=0$. $P({\bf r},t \:\cap \:{\bf r}_0)$ may be expanded as
\[
P({\bf r},t \:\cap \, {\bf r}_0)  =  P({\bf r}_0) P({\bf r},t \: | \: {\bf r}_0) \tag{S13}  \label{Pprod}
\]
where $P({\bf r},t \: | \: {\bf r}_0)$ is the time-dependent conditional probability density function describing a spin pair separated by ${\bf r}$ at time $t$ {\em given} that the same pair was separated by ${\bf r}_0$ at $t=0$. $P({\bf r}_0)$ is the {\it a priori} probability density describing the probability per unit volume of finding a spin pair, equivalent to the reciprocal of the ``volume per spin" equal to $N_H$, the number of spins per unit volume.

The intra-molecular $^1$H--$^1$H distance in water is assumed to be constant so that $|{\bf r}| \!= \!|{\bf r}_0|\! = \!a$, and  $P(r_0) \!= N_H \delta (r_0 – a)$.  Since $\psi=0$ at $t=0$, both $\mathbb{R}^3_0$ and $\mathbb{R}^3$ integrals can be executed, save for the integral involving $\psi$ yielding,
\[
\!\!\!\!\!\!G(t)  \!=\!   \frac{4  \delta N_H }{a^4} \!\!\!\!\int_0^\pi \!\! P_2( \cos \! \psi)  P(\psi,t)  
\sin  \psi \, d \psi. \tag{S14}   \label{eqn:G3App}
\]
A numerical Fourier transformation of $G(t)$ using equations (S9)-(S10) for $P(\psi,t)$ yields the spectral density function $J(\omega)$ and then the relaxation rate $R_1$ as described in the main text. 

In the case of the intra-molecular $^1$H--$^1$H rotor in water, a suitable value for the distance $\delta$ can be obtained.  $\delta$ is the effective thickness of the shell at radius $a$ and appears in the execution of the volume integrals leading to Eq.\,(\ref{eqn:G3App}).  Since $P_2(\cos \! \psi) =1$ at $t\!=\!0$, it is trivial to show that $G(0) = 4  \delta N_H  a^{-4}$ from Eq.\,(\ref{eqn:G3App}). Equation (\ref{eqn:Gt_ave}) gives $G(0) = a^{-6}$ and so 
\[
\delta = \frac{1}{4 N_H a^2} \tag{S15}   \label{eqn:delta}
\]
and  Eq.\,(\ref{eqn:G3App}) reverts to  
\[
\!\!\!\!\!\!G(t)  \!=\!   \frac{1}{a^6} \!\! \int_0^\pi \!\! P_2( \cos \! \psi)  P(\psi,t)  
\sin  \psi \, d \psi. \tag{S16}   \label{eqn:G4App}
\]
The $a^{-6}$ dependence is indicative of a particle-particle dipolar interaction.

\section{Consistency of the L\'{e}vy model for $G(t)$ with alternative models}

The L\'{e}vy model presented here is a more generalised model that incorporates other, simpler, models in the literature.  Previous determinations of the dipolar correlation function $G(t)$ have either assumed Brownian dynamics or restricted consideration to $\alpha \! = \! 1$ (Cauchy-Lorentz dynamics).

A Brownian rotor uses Eqs.\,(S6) and (S8) so that
\[
P(\psi,t) = N(t)  \left[ 1 +2 \sum_{p=1}^\infty e^{-p^2t/\tau} \cos p\psi  \right] \tag{S17}. \label{A9:Pbeta}
\]
Equation\,(\ref{A9:Pbeta}) is substituted into Eq.\,(\ref{eqn:G4App}) and the elementary integrals are executed.  Inserting the normalisation term and expanding the exponentials as a series produces (computer algebra is useful here)
\[
G(t)  =  \frac{8}{15 a^6} \left[ e^{-4t/\tau} + \tfrac{2}{3} e^{-8t/\tau} +\left(\tfrac{2}{3} \right)^2 e^{-12t/\tau} ... \right] . \tag{S18} \label{eqn:GtXpan1}
\]
It is often sufficient to consider just the first two terms of the expansion.  The bi-exponential form for $G(t)$ has been widely used as an improvement to a single-exponential correlation function as a description of field-cycling dispersion curves and in the study of anisotropic dynamics \cite{ropp2001rotational,qvist2012rotational,popov2016mechanism}.
Note that it is the inclusion of the angular boundary conditions $0\!\leq \! \psi \! \leq \pi$ that provide the source of the exponential series expression of Eq.\,(\ref{eqn:GtXpan1}).

\section{Proton-proton separation in a molecule of H$_2$$^{16}$O in the liquid state}

In the following section, justification for the choice of $a$ for the  $^1$H--$^1$H distance in water is made.  This text is adapted from a draft paper by David Sawyer and  Kaz Krynicki (deceased) on translational diffusion in liquid water.

The selection of an accurate value for the proton--proton separation, $r_{\rm HH}$, in the H$_2$$^{16}$O molecule in liquid water is a crucial step in the determination of the intra-molecular contribution to the spin-lattice relaxation rate in water. Even small inaccuracies in its value quickly lead to significant errors when raised to the power six, as in Eq.\,(\ref{eqn:G4App}). 

An examination of the NMR literature on water reveals that the values of $r_{\rm HH}$ employed range from 0.151 to 0.163 nm. Such a variation stems, in part, from the physical state involved, the isotopomer under investigation, the differing techniques used for its determination along with the associated motional averages, for example; $\langle r_{\rm HH}^{-1} \rangle^{-1}$ from scattering experiments, $\langle r_{\rm HH}^{-2} \rangle^{-1/2}$  from rotational spectroscopy and $\langle r_{\rm HH}^{-3} \rangle^{-1/3}$ from NMR \cite{pyper1982,czako2009}. 

The use of $\langle r_{\rm HH}^{-3} \rangle^{-1/3}$ for NMR applications stems of course from the $r^{-3}$ dipolar interaction between spins. Recognition that the $^1$H--$^1$H distance is distributed about a mean immediately highlights that $\langle r_{\rm HH}^{-3} \rangle^{-1/3} \neq \langle r_{\rm HH} \rangle$.   It can be shown that a motional correction to $\langle r_{\rm OH}^{-3} \rangle^{-1/3}$ gives an enhancement of 7.7\% when raised to the sixth power, as compared with an equilibrium value \cite{sawyer.unp}. However, the situation for the unbonded distance $r_{\rm HH}$ is more complex as not only do the vibrational modes, both symmetric and asymmetric, contribute but so does the bending or scissor mode of each OH bond. 

In a detailed {\it ab initio} quantum-mechanical  study of H$_2$$^{16}$O liquid, Czak\'{o} {\it et al} \cite{czako2009} demonstrated that the various motional averages are related through the inequality: mean $\langle r \rangle > \langle r^{-1} \rangle^{-1} >  \langle r^{-2} \rangle^{-1/2}   > \langle r^{-3} \rangle^{-1/3} > r_e$ where $r_e$ represents the equilibrium value. These motional averages in the case of $r_{\rm OH}$ extended over a range of 2\%. The corresponding difference for $r_{\rm HH}$ appears to be more than 3\%. Czak\'{o} also demonstrated that the OH bond shortens on deuteration and that the root-mean-square vibrational amplitude of $r_{\rm HH}$ exceeds that of $r_{\rm OH}$. He also quantified the effect of temperature on bond length, enabling us to say it will be inconsequential for our purposes.
Accordingly, a selection of literature values of $r_{\rm HH}$ have been gathered, mindful of accuracy, which are presented in table I, along with several small adjustments made in the spirit of reference \cite{pyper1982}.

\begin{table}[ht]
	\caption{A log of selected literature values of the intramolecular hydrogen separation in molecules of H$_2$$^{16}$O in the liquid state with adjustments for deuteration and motional averaging. \\[1mm]
	{\small
		$^{\rm (a)}$Values obtained from solutions of LiCl and LiBr in the glass state at 100\,K and relate to interstitial water molecules rather than the molecules of hydration. Effects of libration estimated to be $<$1\% and were ignored by the authors. \\[1mm]
		$^{\rm (b)}$Values obtained from relaxation rate maxima at 200\,K and 200\,MPa where the inter and intramolecular contributions to relaxation were separated. Such pressures and temperatures are believed to have a minimal effect on the value of $r_{\rm HH}$ \cite{czako2009}.}}
	\begin{tabular}{lccccc}\hline \\[-2mm]
		~~~~Source~~~~	& ~~~Technique~~~ &	~~~Experimental~~~   &
		~~~Deuteration~~~   &  ~~Motional average~~  &  ~~Resulting value of~~	 \\
		&  & value &
		adjustment  &    adjustment
		&  $\langle r^{-3} \rangle^{-1/3}$	\\
		&  & (nm)  &
		(\%)  &    
		(\%) &  	(nm)	 \\[1mm] \hline \\[-2mm] 
		Ichikawa {\it et al} \cite{ichikawa1991}		&  Neutron  & $\langle r^{-1}_{\rm D-D} \rangle^{-1} =$  & +0.5  & -0.3  & 0.1573   \\[0.5mm]
		&  scattering & 0.155$\pm$0.001  & &  &  \\[1.5mm]
		Tomberli {\it et al} \cite{tomberli2000}		&  Neutron and  &   $\langle r^{-1}_{\rm H-H} \rangle^{-1} =$ &  n/a &  -0.3   &  01534  \\[0.5mm]
		&  ~~~X-ray scattering~~~ & 0.1539$\pm$0.0002  & &  &  \\[1.5mm]
		Kameda {\it et al} \cite{kameda2018}		&  Neutron & $\langle r^{-1}_{\rm H-H} \rangle^{-1} =$  & n/a  &   -0.3  &   0.1525 \\[0.5mm]
		&  scattering & 0.153$\pm$0.002  & &  &  \\[1.5mm]
		Baianu {\it et al} \cite{baianu1978}		&  NMR & $\langle r^{-3}_{\rm H-H} \rangle^{-1/3} =$  &  n/a &  n/a &  0.155  \\[0.5mm]
		&  lineshape$^{\rm (a)}$ & 0.155$\pm$0.001  & &  &  \\[1.5mm]
		Lang {\it et al} \cite{lang1993}		&  NMR &  $\langle r^{-3}_{\rm H-H} \rangle^{-1/3} =$ &n/a   & n/a  & 0.156   \\[0.5mm]
		&  relaxation$^{\rm (b)}$ & 0.156  & &  &  \\[1.5mm]
		\hline 
	\end{tabular} 
	\label{Table}
\end{table}
\noindent

\bigskip

Based on the above considerations, the mean value of the desired  $\langle r^{-3} \rangle^{-1/3}$ separation for water is found to be 0.1548\,nm with an estimated (SDOM) error of $\pm$0.0005\,nm.

\section{Estimating the relaxation rate uncertainties}

Estimating the relaxation rate uncertainties is challenging due to the complexity of both the dynamics and the process for converting the dynamical time constants to $R_1$.   
Calero {\it et al}\cite{calero20151h}  provide the most relevant  comparison to our work.  
Calero {\it et al} used molecular dynamics (MD) simulation to compute, directly from simulation, the intra- and inter-molecular contributions to $R_1$.  They found $R_1\!=\!0.263$\,s$^{-1}$ with $R_{1,{\rm inter}}/R_1 \!=\! 0.33$ for the TIP3P/2005 potential energy set. 
The bulk diffusion coefficient $D$ was 2.0$\times 10^{-9}$\,m$^2$/s at 298\,K, lower than the experimental value of 2.3$\times 10^{-9}$\,m$^2$/s\cite{krynicki1978pressure} by 13\%.  

The impact on $R_{1,{\rm inter}}$ of a diffusion coefficient can be calculated using the Hwang-Freed formalism\cite{hwang1975dynamic}.  The Hwang-Freed model is a continuum model of diffusion that disallows particles to diffuse through each other. The Hwang-Freed expression yields $R_{1,{\rm inter}}$ via an analytic expression which is dependent on $D$ and a distance-of-nearest-approach parameter $d_{\rm HF}$.    $d_{\rm HF}$  is taken as the value which provides the same $R_{1,{\rm inter}}$ as Calero {\it et al} using first $D\!=\!2.0 \times 10^{-9}$\,m$^2$/s.  It is found that $d_{\rm HF}\!=\!0.140$\,nm.  $d_{\rm HF}$ is now fixed at this value and the diffusion coefficient  increased to $D\!=\!2.3 \times 10^{-9}$\,m$^2$/s.  As a consequence, $R_{1,{\rm inter}}/R_1$ changes from 33\% to 35\% and therefore $R_{1,{\rm intra}}/R_1$ from 67\% to 65\%.  A 13\% increase in the diffusion time constant translates to a 2\% reduction in the fraction of $R_1$ associated with intra-molecular motion.

The intra-molecular contribution to the relaxation rate is more sensitive to dynamical time constants and depends on the interplay of many types of motion.  Our L\'{e}vy model is fit to angular probability density functions generated from MD simulations of water using the flexible SPC/E potential set.  Our estimate of uncertainty requires an evaluation of  the difference between the time constants from MD  and experiment for those dynamics that contribute to the angular rotation of the $^1$H--$^1$H vector.

Toukan and Rahman in 1985\cite{toukan1985molecular} compared MD results using the flexible SPC/E potential set to experiment and  concluded that time constant uncertainties lay in the range $\pm$10--15\% (see Table II).
Praprotnik {\it et al} compared bond stretch, bend and librations with the output of MD simulations also using the flexible SPC/E model.  Ropp and co-workers\cite{ropp2001rotational} undertook both quadrupole NMR measurements and MD simulations of water using the SPC/E potential set for D$_2^{\,\,17}$O enriched water. These authors evaluated the time integral of the second-rank rotational time-correlation function for the O--D bond vector  suggesting that SPC/E  underestimates the time constant by about 20\% at 300\,K. Finally, because molecular tumbling due to collisions also contributes to $R_{1,{\rm intra}}$, the diffusion coefficient obtained from the present MD simulations (1.90$\times 10^{-9}$\,m$^2$/s) is compared with the experimental value of 2.3$\times 10^{-9}$\,m$^2$/s\cite{krynicki1978pressure}. The diffusion {\it time constant} is 17\% too long.

\begin{table}[ht]
	\caption{Estimated uncertainty in dynamical time constants from MD for the flexible SPC/E potential for liquid water.  The percentages represent the deviation of the MD value from  experiment.}
	\begin{tabular}{lcc} \hline \\[-2mm]
		~~~~Source~~~~	& ~~~type of motion ~~~ &	~~~uncertainty in~~~ 	\\
		&  & time constant 	\\ \hline \\[-1mm]
		Toukan and Rahman\cite{toukan1985molecular} & general &  $\pm$10-15\% \\[1.5mm] 
				Praprotnik {\it et al}\cite{praprotnik2004temperature} & stretch &  +1\% \\[1.5mm]
				Praprotnik {\it et al}\cite{praprotnik2004temperature} & bend &  -9\% \\[1.5mm]
				Praprotnik {\it et al}\cite{praprotnik2004temperature} & librations &  -20\% \\[1.5mm]
				Ropp {\it et al}\cite{ropp2001rotational} & O--D bond  &  -20\% \\[1.5mm]
				this work & diffusion &  +17\% \\[1.5mm]		\hline 
	\end{tabular} 
	\label{Table_error}
\end{table}

There are time constants associated with each of many dynamical processes, each important over different time scales.  Some time constants are too long, some too short. We suggest that a combined uncertainty of $\pm 10$\% is reasonable based on limited evidence. $R_{1,{\rm intra}}/R_1$ is therefore re-evaluated with $\tau$ 10\% longer and shorter than the values presented in Fig.\,3 as described in the main article. The result is an estimated uncertainty of $\pm 7$\% to the headline value of $R_{1,{\rm intra}}/R_1$.

\bibliography{References_SM2}